\documentclass[aps,prl,reprint,amsmath,amssymb,nolongbibliography,nofootinbib,
               floatfix,flushbottom,balancelastpage]{revtex4-2}
\usepackage{graphicx,booktabs,bm}
\usepackage{etoolbox}
\makeatletter
\patchcmd{\section}{0.8cm \@plus1ex \@minus .2ex}
  {16pt \@plus2pt \@minus2pt}{}{}
\patchcmd{\section}{0.5cm}{8pt}{}{}
\makeatother

\usepackage[hidelinks]{hyperref}
\hypersetup{
  pdftitle={Next-to-next-to-leading-order QCD predictions for semileptonic Bc to (eta\_c, J/psi) l nu decays at large recoil},
  pdfauthor={Bo-Xuan Shi}
}
\graphicspath{{figures/}}

\newcommand{\msbar}{\overline{\mathrm{MS}}}

\newcommand{\GeV}{\,\mathrm{GeV}}
\newcommand{\GeVsq}{\,\mathrm{GeV}^2}
\newcommand{\dd}{\mathrm{d}}

\begin{document}

\title{\texorpdfstring{Next-to-next-to-leading-order QCD predictions for semileptonic\\\(B_c\to\{\eta_c,J/\psi\}\ell\nu\) decays at large recoil}{Next-to-next-to-leading-order QCD predictions for semileptonic Bc to (eta\_c, J/psi) l nu decays at large recoil}}
\author{Bo-Xuan Shi}
\email{shibx@pku.edu.cn}
\affiliation{Center for High Energy Physics, Peking University, Beijing 100871, China}

\begin{abstract}
We present next-to-next-to-leading-order (NNLO) QCD predictions for
\mbox{\(B_c\to\eta_c\ell\nu\)} and \mbox{\(B_c\to J/\psi\ell\nu\)} at large recoil.
We compute the form factors through two loops at finite charm-to-bottom
mass ratio and leading order in the heavy-quark relative velocities. By applying the
modern effective field theory formalism, we carry out, for the first
time in these transitions, explicit two-loop matching onto physical and
evanescent operators in nonrelativistic QCD (NRQCD), including the finite
terms from evanescent-to-physical mixing. The cancellation of residual
infrared poles verifies NRQCD factorization at representative recoils.
At the reference inputs, the NNLO corrections raise the leading-velocity
hard-exchange predictions for the Standard Model partial widths in
\mbox{\(m_\tau^2\le q^2\le5\GeVsq\)} by about \(30\%\) relative to
next-to-leading order in both channels and reduce their sensitivity to the
renormalization scale. We also determine the
binned lepton-flavor ratios and find that the \(J/\psi\) helicity-interference
moment shifts toward the lattice-QCD result.
\end{abstract}

\maketitle

\setlength{\abovedisplayskip}{6pt plus 2pt minus 3pt}
\setlength{\belowdisplayskip}{\abovedisplayskip}
\setlength{\abovedisplayshortskip}{\abovedisplayskip}
\setlength{\belowdisplayshortskip}{\abovedisplayskip}
\setlength{\textfloatsep}{14pt plus 2pt minus 3pt}
\setlength{\dbltextfloatsep}{14pt plus 2pt minus 3pt}
\setlength{\parskip}{0pt plus 0.5pt}

\section{Introduction}
\label{sec:intro}

The semileptonic decays \(B_c\to\eta_c\ell\nu\) and
\(B_c\to J/\psi\ell\nu\) probe the same \(b\to c\) charged current
with a charm antiquark spectator. Measurements of the full-phase-space
\(\tau\)-to-\(\mu\) branching-fraction ratio \(R(J/\psi)\) test lepton
flavor universality in the Standard Model
(SM)~\cite{LHCb2017,LHCb2026,CMS2024,CMS2025}.
The angular distribution of the semileptonic cascade with
\(J/\psi\to\mu^+\mu^-\) also probes helicity interference~\cite{HPQCDLFU2020}.
Lattice QCD provides the \(B_c\to J/\psi\) form factors over the full
recoil range~\cite{HPQCD2020,HPQCD2025}.

Away from zero recoil, the spectator acquires its recoil through
hard gluon exchange at leading order in the relative velocities. In
NRQCD~\cite{CaswellLepage,BBL}, each form factor then factorizes into
a short-distance coefficient and a local \(S\)-wave matrix element for
each meson~\cite{QiaoSunYuan2012,QiaoZhu}.
A QCD sum-rule analysis at spacelike momentum transfer supports this
mechanism for \(B_c\to J/\psi\) near maximum recoil~\cite{BordoneSumRules}.
The gluon virtuality vanishes at zero recoil, so the hard-exchange
description breaks down near that endpoint.

The sizable one-loop corrections at finite mass
ratio~\cite{QiaoSunYuan2012,QiaoZhu,TaoTensor2022} motivate a two-loop
calculation. In the hierarchy \(m_b\gg m_c\), the one-loop form
factors~\cite{BellThesis} and the soft-overlap contribution to
\(B_c\to\eta_c\) at double-logarithmic accuracy~\cite{BellSoftOverlap,BellSCET}
are known. The leading \(S\)-wave NRQCD currents acquire anomalous
dimensions at two loops~\cite{CzarneckiMelnikovThreshold,OnishchenkoVeretin,ChenQiaoLeptonic,HoangRuiz,Pan}.
Their renormalization predicts the infrared pole of the ultraviolet-renormalized QCD
amplitude and thus provides a two-loop test of factorization.

By applying the modern effective field theory formalism, we perform
the first explicit two-loop matching of these transitions onto a basis
of physical and evanescent NRQCD operators at leading order in the
velocity expansion and at finite charm-to-bottom mass ratio. The NNLO
form factors retain the finite terms from \mbox{evanescent-to-physical}
mixing.

At large recoil we determine the differential and partial widths and
the binned lepton-flavor ratios, and study \(J/\psi\) helicity interference.

\section{Form factors and factorization}
\label{sec:factorization}

For \(B_c^-(p)\to H(p')\) with \(H=\eta_c,J/\psi\), let
\(q=p-p'\), \(M^2=p^2\), and \(m^2=p'^2\).

For \(\eta_c\), we use the vector decomposition of
Ref.~\cite{MurphySoni2018} and the tensor convention of
Ref.~\cite{TranIvanovKornerSantorelli}, with
\(\sigma^{\mu\nu}=i[\gamma^\mu,\gamma^\nu]/2\),
\begin{align}
 &\langle\eta_c|\bar c\gamma^\mu b|B_c\rangle
 =f_+(q^2)\left[(p+p')^\mu-\frac{M^2-m^2}{q^2}q^\mu\right]
 \nonumber\\
 &\qquad+f_0(q^2)\frac{M^2-m^2}{q^2}q^\mu ,
 \nonumber\\
 &\langle\eta_c|\bar c\sigma^{\mu\nu}b|B_c\rangle
 =-\frac{2i f_T(q^2;\mu_T)}{M+m}
     (p^\mu p'^\nu-p^\nu p'^\mu),
 \label{eq:etac-form-factors}
\end{align}
where \(\mu_T\) is the \(\msbar\) renormalization scale of the tensor
current~\cite{GraceyTensor}.

For polarization \(\xi\) with \(p'\cdot\xi=0\), the \(J/\psi\) form
factors are defined with \(\varepsilon_{0123}=+1\)~\cite{QiaoZhu,HPQCD2020} as
\begin{align}
 &\langle J/\psi|\bar c\gamma^\mu b|B_c\rangle
 =\frac{2i V}{M+m}\,
        \varepsilon^{\mu\nu\rho\sigma}\xi_\nu^*p'_\rho p_\sigma ,
        \nonumber\\
 &\langle J/\psi|\bar c\gamma^\mu\gamma_5 b|B_c\rangle\nonumber\\
 &\quad=(M+m)A_1\left(\xi^{*\mu}-\frac{\xi^*\cdot q}{q^2}q^\mu\right)
        \nonumber\\[-1mm]
 &\qquad-\frac{\xi^*\cdot q}{M+m}A_2
      \left(p^\mu+p'^\mu-\frac{M^2-m^2}{q^2}q^\mu\right)
        \nonumber\\
 &\qquad+2m A_0\frac{\xi^*\cdot q}{q^2}q^\mu .
        \label{eq:jpsi-form-factors}
\end{align}

At leading order in the relative velocities, the meson masses reduce to
their partonic values \(M=m_b+m_c\) and \(m=2m_c\), with momenta
\(p=Mv\) and \(p'=mv'\). In terms of the mass ratio \(z=m_c/m_b\) and
recoil \(w=v\cdot v'\),
\begin{equation}
 q^2/m_b^2=(1+z)^2+4z^2-4z(1+z)w .
 \label{eq:kinematics}
\end{equation}

The spectator receives momentum \(k_g=m_c(v'-v)\) through a Born
gluon with \(Q_g=\sqrt{-k_g^2}=m_c\sqrt{2(w-1)}\).
At fixed \(w>1\), the bound-state momentum
\(p_{\rm rel}\sim m_c v_{\rm rel}\) satisfies
\(p_{\rm rel}/Q_g\sim v_{\rm rel}/\sqrt{2(w-1)}\ll1\)
in the velocity expansion~\cite{BBL}.
The form factors \(F_i\) obey the
leading-velocity factorization formula
\begin{align}
 F_i&=C_i(w,z,m_b;\mu_R,\mu_\Lambda)\,
       \Phi_{B_c}(\mu_\Lambda)\Phi_H^*(\mu_\Lambda)
       + O(v_{\rm rel}^2),\nonumber\\
 C_i&=C_i^{\mathrm{Born}}\!
      \left[1+a_s c_i^{(1)}+a_s^2 c_i^{(2)}+O(a_s^3)\right].
 \label{eq:factorization}
\end{align}
We use the \(\msbar\) coupling \(a_s=\alpha_s(\mu_R)/(4\pi)\),
with \(\mu_\Lambda\) the NRQCD factorization scale.
Here \(c_i^{(\ell)}\) is the \(\ell\)-loop hard coefficient normalized
to the single-gluon exchange \(C_i^{\rm Born}\propto\alpha_s\).
The coefficient for \(f_T\) also depends on \(\mu_T\).
\par

The bound-state factors are vacuum-to-meson matrix elements of local
color-singlet currents. With \(\psi_Q\) and \(\chi_Q^\dagger\) the
Pauli fields that annihilate a quark and an antiquark of flavor
\(Q\)~\cite{BBL}, we define
\begin{equation}
 \begin{aligned}[b]
 \Phi_{B_c}(\mu_\Lambda)&=\langle0|[\chi_c^\dagger\psi_b]_{\mu_\Lambda}|B_c\rangle,\\
 \Phi_{\eta_c}(\mu_\Lambda)&=\langle0|[\chi_c^\dagger\psi_c]_{\mu_\Lambda}|\eta_c\rangle,\\
 \Phi_{J/\psi}(\mu_\Lambda)\,\xi^i
 &=\langle0|[\chi_c^\dagger\sigma^i\psi_c]_{\mu_\Lambda}|J/\psi(\xi)\rangle .
 \end{aligned}
 \label{eq:meson-currents}
\end{equation}
The identity and \(\sigma^i\) select the \({}^1S_0\) and \({}^3S_1\)
channels, respectively, in the meson rest frames.
With nonrelativistic state normalization, each meson
matrix element is \(\Phi=\sqrt{2N_c}\,\psi(0)\) at this order,
where \(N_c=3\) and \(\psi(0)\) is its wave function at the origin.

\section{Matching and infrared subtraction}
\label{sec:matching}

To extract the hard coefficients, we calculate the on-shell QCD amplitude
\begin{equation}
 {\cal A}_\Gamma =
 \langle c(m_cv')\,\bar c(m_cv')|
       \bar c(0)\Gamma b(0)
 |b(m_bv)\,\bar c(m_cv)\rangle .
 \label{eq:qcd-correlator}
\end{equation}
We keep the external spin indices open in \(D=4-2\epsilon\) and
expand the amplitude in partonic matrix elements of physical and
evanescent NRQCD operators.

To represent the meson spin channels in naive dimensional
regularization (NDR), we use \(\Sigma^i=\gamma^i\gamma_5\) on the
positive-energy subspace in each pair rest frame. These matrices
satisfy \(\{\Sigma^i,\Sigma^j\}=2\delta^{ij}\) in \(D-1\) spatial
dimensions~\cite{HoangRuiz}. For an incoming spin singlet, the operators contributing to the physical meson transitions are
\begin{equation}
 P_{\eta_c}=(\psi_c^\dagger\chi_c)(\chi_c^\dagger\psi_b),\quad
 P_{J/\psi}^i=(\psi_c^\dagger\Sigma^i\chi_c)(\chi_c^\dagger\psi_b).
 \label{eq:physical-operators}
\end{equation}

In four dimensions a Fierz rearrangement converts crossed QCD chains
into NRQCD bilinears. Its \(D\)-dimensional remainder is
evanescent~\cite{DuganGrinstein,HerrlichNierste}.
Antisymmetrized products with three or more indices transverse to both
\(v\) and \(v'\) also vanish in four dimensions. Both kinds of evanescent
structures enter the operator basis used to construct the infrared
subtraction, as in calculations of exclusive form factors and
light-cone operators~\cite{WangShenPion,GaoPion,JiPionEM,HuangNucleonFF,ShiSigmaFF,HuangNucleonDA}.

A Fierz remainder occurs already in transverse Born gluon exchange
for the contracted vector current \(v_\mu\bar c\gamma^\mu b\),
\begin{equation}
 \begin{aligned}[b]
 E_\perp
 &=\sum_a
   (\psi_c^\dagger\Sigma^\parallel\Sigma^a\psi_b)
   (\chi_c^\dagger\Sigma^a\Sigma^\parallel\chi_c)\\
 &\quad+P_{\eta_c}
    -(\psi_c^\dagger\Sigma^\parallel\chi_c)
     (\chi_c^\dagger\Sigma^\parallel\psi_b).
 \end{aligned}
 \label{eq:evanescent-line}
\end{equation}
Here \(\parallel\) denotes the recoil axis, oriented along the outgoing
meson three-momentum in the \(B_c\) rest frame, and \(a\) runs over the
\(D-2\) common transverse directions.
The longitudinal triplet--triplet term is required by the Fierz
identity with open spin indices and drops out upon projection onto
the incoming spin singlet.

\begin{samepage}
Let \(\langle\mathcal O_j\rangle^{(0)}\) be the tree matrix elements
of the physical and evanescent operators in the states of
Eq.~\eqref{eq:qcd-correlator}. After ultraviolet (UV) renormalization,
the amplitude satisfies
\begin{equation}
 {\cal A}_\Gamma
 =\sum_j A_{\Gamma,j}\,\langle{\cal O}_j\rangle^{(0)}
 =\sum_{k,j} C_{\Gamma,k} Z_{kj}\,\langle{\cal O}_j\rangle^{(0)}.
 \label{eq:operator-matching}
\end{equation}
\par
\end{samepage}
The \(C_{\Gamma,k}\) multiply renormalized NRQCD operators.
At zero relative momentum, bare NRQCD loops are scaleless, so the
renormalized matrix element of operator \(k\) is a sum of tree structures \(j\)
weighted by \(Z_{kj}=\delta_{kj}+a_s^2Z_{kj}^{(2)}+O(a_s^3)\).

We expand the QCD coefficients as
\begin{equation}
 A_{\Gamma,j}
 ={\textstyle\sum\limits_{\ell=0}^{2}}(Z_\alpha a_s)^{\ell+1}A_{\Gamma,j}^{(\ell)}
 +{\textstyle\sum\limits_{\ell=1}^{2}}a_s^{\ell+1}\delta A_{\Gamma,j}^{\mathrm{ct},(\ell)}
 +O(a_s^4),
 \label{eq:qcd-coefficient-expansion}
\end{equation}
here \(A_{\Gamma,j}^{(\ell)}\) is the bare \(\ell\)-loop coefficient,
\(Z_\alpha\) renormalizes the coupling, and
\(\delta A_{\Gamma,j}^{\mathrm{ct},(\ell)}\) collects terms with
mass, external-field, or current counterterm insertions.
We renormalize masses and fields on
shell~\cite{BroadhurstGraySchilcher,BekavacOnShell} and differentiate
internal masses at fixed external kinematics.
Writing \(C_{\Gamma,j}^{(\ell)}\) for the coefficient of
\(a_s^{\ell+1}\) in \(C_{\Gamma,j}\), Eq.~\eqref{eq:operator-matching} gives
\begin{align}
 C_{\Gamma,j}^{(1)}
 &=A_{\Gamma,j}^{(1)}+Z_\alpha^{(1)} A_{\Gamma,j}^{(0)}
     +\delta A_{\Gamma,j}^{\mathrm{ct},(1)},\nonumber\\
 C_{\Gamma,j}^{(2)}
 &=A_{\Gamma,j}^{(2)}+2Z_\alpha^{(1)} A_{\Gamma,j}^{(1)}
     +Z_\alpha^{(2)} A_{\Gamma,j}^{(0)}\nonumber\\
 &\quad+\delta A_{\Gamma,j}^{\mathrm{ct},(2)}
     -\sum_k A_{\Gamma,k}^{(0)} Z_{kj}^{(2)}.
 \label{eq:matching-coefficients}
\end{align}
Since \(Z_{kj}^{(1)}=0\), the last term acts on the full Born decomposition,
so only evanescent operators present there can contribute to the finite
physical coefficients at NNLO. At fixed recoil and leading
order in the relative velocities, eikonal couplings cancel within each
color singlet~\cite{BodwinGarciaLee}, and in the threshold
expansion~\cite{BenekeSmirnov,Jantzen} potential pinches are confined to
each pair. The physical pole in \(Z_{kj}\) therefore follows from the
inverse product of the two \(S\)-wave current renormalization factors
\(Z_S\), defined by \(\Phi^{\rm bare}=Z_S\Phi\) for spin
\(S=0,1\)~\cite{OnishchenkoVeretin,HoangRuiz,Pan}.

The current counterterms are diagonal on antisymmetric spin tensors.
Their spin-dependent part is proportional to \(n(D-1-n)\) for a
product of \(n\) antisymmetrized spin matrices~\cite{HoangRuiz}, with
normalization fixed by the singlet and triplet counterterms.
We apply the counterterms in this spin basis and reexpress the result
with the same Fierz remainders, keeping their four-dimensional Pauli
coefficients fixed. The \(O(\epsilon)\) spin terms multiply the pole and
give a finite evanescent-to-physical subtraction. For the Born
structure in Eq.~\eqref{eq:evanescent-line} it is governed by
\begin{equation}
 Z_{E_\perp P_{\eta_c}}^{(2)}
 =\frac{16\pi^2 C_F^2}{3}
   \left[\frac{z}{(1+z)^2}+\frac14\right].
 \label{eq:born-ep-mixing}
\end{equation}
Here \(C_F=(N_c^2-1)/(2N_c)\). Section~S2 lists the complete set of
Born evanescent operators and their finite mixing coefficients.

\section{Two-loop calculation}
\label{sec:calculation}

\begin{figure}[!tb]
 \centering
 \setlength{\belowcaptionskip}{0pt}
 \includegraphics[width=\columnwidth,trim={2.99bp 4.07bp 3.00bp 3.61bp},clip]{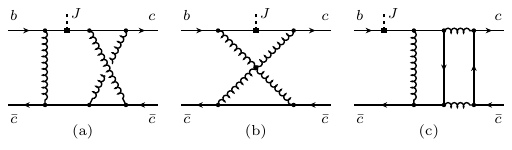}
 \caption{Representative two-loop diagrams. (a) Crossed gluon exchange.
 (b) Four-gluon vertex. (c) Two-gluon production of the final
 \(c\bar c\) pair, with an additional exchange in the incoming pair.
 Squares mark the weak-current insertion \(J\), and dots mark QCD vertices.}
 \label{fig:diagrams}
\end{figure}

We use QGRAF~\cite{Nogueira} to generate \(1{,}610\) two-loop diagrams,
keeping finite charm and bottom masses in closed quark loops.
Two-gluon production of the final \(c\bar c\) pair gives an absorptive
part for \(\eta_c\) starting at one loop. Figure~\ref{fig:diagrams}(c)
shows a two-loop example. Charge conjugation requires at least three
gluons to produce a color-singlet \(J/\psi\) in this class of diagrams,
so its absorptive contribution starts at two loops.

We use \textsc{Form}~\cite{FORM5} to extract the bare coefficients
\(A_{\Gamma,j}^{(\ell)}\) in the \(D\)-dimensional operator basis of
Eq.~\eqref{eq:operator-matching}. We reduce the resulting scalar
loop integrals using integration-by-parts
identities~\cite{TkachovIBP,ChetyrkinTkachov} and the Laporta
algorithm~\cite{Laporta} in \textsc{Kira}~\cite{Kira,Kira2,Kira3}.
The reduced \(\eta_c\) and \(J/\psi\) amplitudes involve \(1{,}472\) and \(1{,}236\)
master integrals, respectively, with coefficients rational in \(D,w,z\).

To evaluate these integrals across the recoil range, we use
\(x=2v\cdot p'/m_b=4zw\), the charmonium energy in the \(B_c\) rest
frame in units of \(m_b/2\). We compute the Laurent coefficients at the
common boundary point \((x,z)=(7/5,6/19)\) using auxiliary mass
flow~\cite{LiuMaWangAMF,LiuMaAMF,LiuMaIteratedAMF} in
\textsc{AMFlow}~\cite{AMFlow,AMFlow2}, direct Feynman-parameter
integration with sector decomposition~\cite{BinothHeinrich,KanekoUeda} in
\textsc{pySecDec}~\cite{pySecDec}, and analytic evaluation for simpler
integrals. This use of mass flow for kinematic boundary conditions
follows Ref.~\cite{LiuMaTaoZhangAMF}. We evolve the boundary values
with kinematic differential equations~\cite{KotikovDE,RemiddiDE,GehrmannRemiddiDE}
\begin{equation}
 \partial_j\boldsymbol I=\mathsf K_j(D,x,z)\boldsymbol I,\qquad j=x,z ,
 \label{eq:master-de}
\end{equation}
where \(\boldsymbol I\) includes auxiliary integrals needed
to close the system.

We solve these equations by local series along complex contours on
the Feynman sheet~\cite{MorielloDE,ChenDoubleCharmonium}. Evolution in
\(x\) supplies the recoil grids for both channels, while evolution in
\(z\) gives the coefficients at the additional \(J/\psi\) mass pairs.
Direct evaluations at selected recoils check the continuation.

We test Eq.~\eqref{eq:factorization} at two loops using the single
infrared pole left after QCD UV renormalization. At order \(a_s^2\)
relative to Born, the NRQCD current counterterms predict a residue of
\(4[\gamma_0^{(2)}(z)+\gamma_{S_H}^{(2)}(1)]\), where
\(\gamma_S^{(2)}(r)\) is the coefficient of \((\alpha_s/\pi)^2\) in
\(\dd\ln Z_S/\dd\ln\mu_\Lambda\) for a quark pair of mass ratio
\(r\)~\cite{Pan}. The final spin \(S_H\) is \(0\) for \(\eta_c\)
and \(1\) for \(J/\psi\). At the reference mass ratio,
we confirm pole cancellation numerically at all 11 positive-\(q^2\)
grid points for \(\eta_c\) and at 6 for \(J/\psi\). The associated
factorization-scale logarithms compensate the running of the two matrix
elements through NNLO. After infrared subtraction in the operator
basis, we take the four-dimensional limit and project onto the meson
spin states. Dividing by the corresponding Born form factors gives
\(c_i^{(2)}\).

For a spin-singlet matrix element with quark mass ratio \(r\), the
two-loop current counterterm~\cite{HoangRuiz} gives the finite conversion
between NDR and Larin's prescription~\cite{Larin},
\begin{equation}
 \Phi_{\mathrm{NDR}}=
 \left[1-a_s^2\frac{32\pi^2 C_F^2 r}{(1+r)^2}
      +O(a_s^3)\right]\Phi_{\mathrm{Larin}}.
 \label{eq:spin-conversion}
\end{equation}
Factorization fixes the inverse conversion of the hard coefficients.
Using common scalar integrals, independent NDR and Larin numerator
calculations reproduce this shift in both channels at \(x=7/5\) and
\(99/70\) for \(z=6/19\).

\begin{table}[!b]
\centering
\caption{Born-normalized NDR coefficients \(c_i^{(\ell)}\) of
Eq.~\eqref{eq:factorization} at \(q^2=5\GeVsq\), \(z=6/19\),
\(m_b=4.75\GeV\), and \(\mu_R=\mu_\Lambda=\mu_T=m_b\).}
\label{tab:benchmark}
\begin{tabular}{@{}lrr@{}}
\toprule
 \(F_i\) & \multicolumn{1}{c}{\(c_i^{(1)}\)} & \multicolumn{1}{c}{\(c_i^{(2)}\)} \\
\midrule
 \(f_+\) & \(48.9223+0.3547i\) & \(193.5721+37.4463i\) \\
 \(f_0\) & \(48.2570+0.3217i\) & \(148.7894+35.9459i\) \\
 \(f_T\) & \(56.6634+0.3800i\) & \(735.7005+39.8537i\) \\
 \addlinespace
 \(V\) & \(48.0770\phantom{{}+0.0000i}\) & \(621.7453+0.1707i\) \\
 \(A_0\) & \(45.3795\phantom{{}+0.0000i}\) & \(372.4804+0.0921i\) \\
 \(A_1\) & \(42.9010\phantom{{}+0.0000i}\) & \(225.6271+0.1513i\) \\
 \(A_2\) & \(34.4591\phantom{{}+0.0000i}\) & \(-231.1827+0.3761i\) \\
\bottomrule
\end{tabular}
\end{table}

At maximum recoil, regularity of Eqs.~\eqref{eq:etac-form-factors}
and \eqref{eq:jpsi-form-factors} gives \(f_0=f_+\) and
\(2mA_0=(M+m)A_1-(M-m)A_2\) at \(q^2=0\). We check these relations
by extracting \(f_0\) and \(A_0\) from the scalar and pseudoscalar
matrix elements using the Ward identities in the pole-mass scheme.\par

\section{Form-factor results}
\label{sec:results}

We evaluate the hard coefficients at the reference pole masses
\(m_b=4.75\GeV\) and \(m_c=1.50\GeV\)~\cite{TaoTensor2022}.
Table~\ref{tab:benchmark} lists the one- and two-loop coefficients for
both channels at \(q^2=5\GeVsq\). The ancillary
file gives the NDR coefficients at these masses on grids spanning
\(0\le q^2\le6.5\GeVsq\) for \(\eta_c\) and \(0\le q^2\le8\GeVsq\)
for \(J/\psi\).\par

\begin{figure*}[!t]
\begin{minipage}[t]{\columnwidth}
 \centering
 \setlength{\belowcaptionskip}{0pt}
 \setlength{\abovecaptionskip}{5pt}
 \includegraphics[width=\columnwidth,trim={15.92bp 9.29bp 15.97bp 2.26bp},clip]{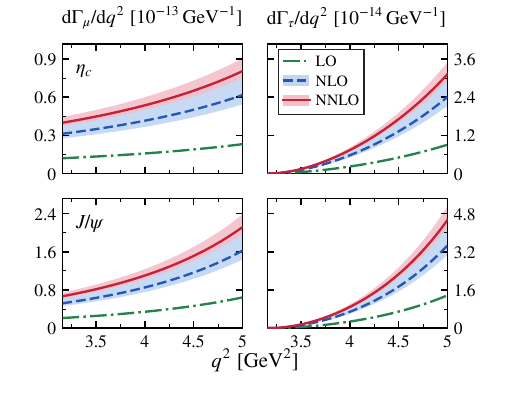}
 \caption{Leading-velocity SM differential widths for
 \(B_c\to\eta_c\ell\nu\) (top) and \(B_c\to J/\psi\ell\nu\) (bottom)
 over \(m_\tau^2\le q^2\le5\GeVsq\), for \(\ell=\mu\) (left)
 and \(\ell=\tau\) (right).
 The central curves use \(\mu_R=\mu_\Lambda=m_b\).
 The bands show how the NLO and NNLO curves vary with
 \(\mu_R\) between \(3\) and \(6\GeV\) at fixed \(\mu_\Lambda\).}
 \label{fig:differential-width-scale}
\end{minipage}
\hfill
\begin{minipage}[t]{\columnwidth}
 \centering
 \setlength{\belowcaptionskip}{0pt}
 \setlength{\abovecaptionskip}{5pt}
 \includegraphics[width=\columnwidth,trim={2.33bp 4.07bp 5.21bp 1.75bp},clip]{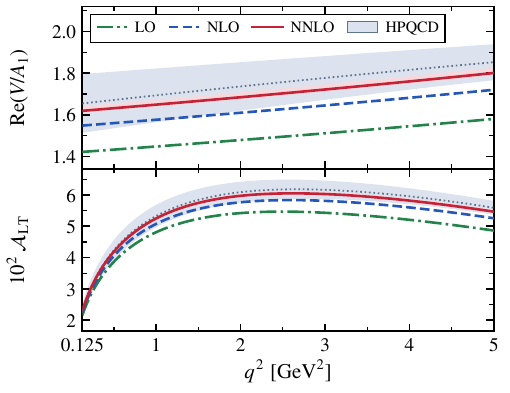}
 \caption{The \(B_c\to J/\psi\) form-factor ratio
 \(\operatorname{Re}(V/A_1)\) (upper) and the semimuonic angular moment
 \(\mathcal A_{\rm LT}\) from Eq.~\eqref{eq:alt} (lower) at \(\mu_R=m_b\).
 The pink bands show the effect of varying \(\mu_R\) from \(3\)
 to \(6\GeV\) at fixed masses.
 Dotted curves and gray-blue bands show the central values and
 \(1\sigma\) intervals reconstructed from the correlated HPQCD
 continuum fit~\cite{HPQCD2025}.}
 \label{fig:va1}
\end{minipage}
\end{figure*}

\begin{samepage}
\section{Phenomenology at large recoil}
\label{sec:fiducial}

Near zero recoil, the momentum transfer to the
spectator no longer provides a hard scale for the velocity expansion.
We compare radiative corrections in the two channels over the common
window \(q^2\le5\GeVsq\).
We use partonic kinematics for the hard coefficients and the central
values of the meson and lepton masses~\cite{PDG2026} for the helicity
amplitudes and decay phase space. Evolving
\(\alpha_s(m_Z)=0.1180\)~\cite{PDG2026} with the five-flavor two-loop
beta function~\cite{CaswellBeta,JonesBeta} gives
\(\alpha_s(m_b)\simeq0.215408\).
With hadron masses, we preserve the axial endpoint relation by adding
\(\tfrac12[M/m-(m_b+m_c)/(2m_c)](A_1-A_2)\) to \(A_0\) throughout
the recoil window, a kinematic correction beyond the leading-velocity
approximation. We retain the complex helicity amplitudes and
expand ratios through NNLO, integrating numerator and denominator
separately for recoil bins.\par
\end{samepage}

For the absolute rates, we use \(G_F=1.16637\times10^{-5}\,\mathrm{GeV}^{-2}\)
and \(\lvert V_{cb}\rvert=0.0406\)~\cite{ChenQiaoLeptonic}.
We take \(\lvert\psi_{B_c}(0)\rvert=0.3615\GeV^{3/2}\) and
\(\lvert\psi_{\eta_c}(0)\rvert=\lvert\psi_{J/\psi}(0)\rvert
=0.283\GeV^{3/2}\) from Ref.~\cite{QiaoZhu} and use them as NDR inputs
at \(\mu_\Lambda=m_b\).

Figure~\ref{fig:differential-width-scale} shows
that NNLO raises the differential widths and narrows the scale bands
above the respective lepton thresholds in the common recoil window.
At \(\mu_R=m_b\),
the SM partial widths increase by about \(30\%\)
from NLO to NNLO in both channels and both lepton modes. The squared one-loop amplitude
supplies most of the increase. Born--two-loop interference contributes
about \(7\%\) of the NLO width for \(J/\psi\) and \(4\%\) for
\(\eta_c\).
For each partial width, we take the difference between its largest
and smallest values as \(\mu_R\) varies from \(3\) to
\(6\GeV\)~\cite{QiaoZhu} and divide by its value at \(\mu_R=m_b\)
at the same order. The relative scale variation of the leading-velocity
widths is about half as large at NNLO as at NLO.

For the central inputs, the NNLO hard-exchange spectrum for
\(B_c\to J/\psi\mu\nu\) lies well above the spectrum reconstructed
from the HPQCD form factors~\cite{HPQCD2025} in the
common bin \(I=[m_\tau^2,5\GeVsq]\). Both the excess and the steeper
\(q^2\) dependence are already present at Born level. The recoil
dependence is consistent with the enhancement of the Born kernel as
the spectator-gluon virtuality decreases in magnitude~\cite{QiaoZhu}.

The rate enhancement largely cancels in the binned ratios
\(R_I(H)=\Gamma_I(B_c\to H\tau\nu)/\Gamma_I(B_c\to H\mu\nu)\).
From NLO to NNLO, \(R_I(\eta_c)\) changes from \(0.1916\) to \(0.1901\),
and \(R_I(J/\psi)\) from \(0.1177\) to \(0.1174\).

The different radiative corrections to \(V\) and \(A_1\)
(Table~\ref{tab:benchmark}) affect the angular correlations in the
cascade with \(J/\psi\to\mu^+\mu^-\)~\cite{HPQCDLFU2020}.
The form factor \(V\) enters the transverse helicity amplitudes
with opposite signs,
\begin{equation}
 \begin{aligned}[b]
 H_\pm&=(M+m)A_1\mp\frac{\sqrt{\lambda}}{M+m}V,\\
 \lambda&=[(M+m)^2-q^2][(M-m)^2-q^2].
 \end{aligned}
 \label{eq:helicity}
\end{equation}
The helicity difference is
\(|H_-|^2-|H_+|^2=4\sqrt{\lambda}\operatorname{Re}(VA_1^*)\).
The ratio \(\operatorname{Re}(V/A_1)\) measures this interference
relative to \(|A_1|^2\).

\begin{samepage}
We probe interference with the longitudinal amplitude through the
normalized angular moment
\begin{equation}
 \mathcal A_{\rm LT}
 =-\big\langle\sin\theta_W\sin(2\theta_{J/\psi})\cos\chi\big\rangle_{q^2}.
 \label{eq:alt}
\end{equation}
\par
\end{samepage}
The brackets denote a rate-weighted average at fixed \(q^2\).
In the massless limit for the muon from the \(B_c\) decay, the angular
weight selects \(\operatorname{Re}[(H_--H_+)H_0^*]\), with \(H_0\) the
longitudinal amplitude. For the numerical results we retain this muon
mass using the angular distribution of Ref.~\cite{HPQCDLFU2020}.
The masses of the two muons from \(J/\psi\) decay give a factor
\(2\beta_\psi^2/(3-\beta_\psi^2)\), where \(\beta_\psi^2=1-4m_\mu^2/m^2\).

NNLO shifts both curves in Fig.~\ref{fig:va1} toward the HPQCD
central values. In the bin \(q^2\in[0.125,5]\GeVsq\), the angular
moment increases by about \(4\%\), with both predictions within the
lattice \(1\sigma\) interval. Finite two-loop matching drives the
increase, partly offset by terms fixed at one loop. Mass-input
variations and relative-velocity corrections~\cite{ZhuRelativistic}
can each produce shifts comparable to the NNLO effect.
In the common \(\tau\)-threshold bin, the \(J/\psi\) longitudinal
fraction \(\Gamma_L/\Gamma\)~\cite{HPQCDLFU2020} shifts by about
\(10^{-3}\) in either lepton mode.

The tensor form factor \(f_T\) enters the interference of a left-handed
tensor interaction with the SM amplitude in
\(B_c\to\eta_c\tau\nu\)~\cite{TranIvanovKornerSantorelli}.
In the common bin at the reference scales, the tensor--SM interference
per unit real tensor coupling, normalized to the SM partial width,
increases by \(4.8\%\) from NLO to NNLO.

\section{Conclusions}
\label{sec:conclusion}

We have obtained the NNLO form factors for \(B_c\to\eta_c,J/\psi\)
at leading order in the heavy-quark relative velocities. By applying the modern effective
field theory formalism, we have carried out the first explicit
two-loop matching of these transitions onto physical and evanescent
NRQCD operators at finite mass ratio.
The matching retains the finite evanescent-to-physical contributions.
The cancellation of the remaining infrared poles verifies two-loop
NRQCD factorization at the tested recoils. The NDR and Larin results also agree after finite
conversion at two kinematic points.

\begin{samepage}
At large recoil, NNLO corrections substantially enhance the
leading-velocity partial widths and reduce their relative variation
under changes in \(\mu_R\) by about a factor of two, while leaving the
\(\tau\)-to-\(\mu\) ratios and the \(J/\psi\) longitudinal polarization
fraction nearly unchanged.
The \(J/\psi\) helicity-interference moment moves closer to the
lattice-QCD result, while the absolute spectrum differs in normalization
and recoil shape~\cite{HPQCD2025}.\par
\end{samepage}

\section*{Note added}

While this manuscript was being finalized, Tao \emph{et al.} reported
an independent NNLO calculation~\cite{TaoEtAl2026NNLO}. With the
wave-function definition of their Eq.~(4.4), the direct tree-level
check in Sec.~S1 gives \(V\) and \(f_+\) a factor \(\pi\) larger than
their ancillary values, corresponding to a factor \(\pi^2\) in the
Born rate normalization. At common
kinematics and with the coupling, tensor-current, and NRQCD conventions
aligned, the Born-normalized two-loop coefficients agree in
\(B_c\to J/\psi\), and all three coefficients differ in \(B_c\to\eta_c\).
This discrepancy
could arise from an inconsistent treatment of \(\gamma_5\) in
the covariant trace formalism.\par

\section*{Acknowledgments}

The author thanks Yu-Ming Wang for helpful discussions and acknowledges
support from the National Natural Science Foundation of China under
Grant No.~12325503.\par

\section*{Data availability}

The complex NDR matching coefficients are available in
\texttt{anc/form-factors.wl}.\par

\renewcommand{\bibfont}{\fontsize{9}{10.2}\selectfont\interlinepenalty=10000\relax}
\setlength{\bibsep}{0pt plus 0.6pt}
\bibliography{references}

@article{LHCb2017,
  author = {Aaij, Roel and others},
  collaboration = {LHCb},
  title = {Measurement of the ratio of branching fractions {\(\mathcal B(B_c^+\to J/\psi\tau^+\nu_\tau)/\mathcal B(B_c^+\to J/\psi\mu^+\nu_\mu)\)}},
  journal = {Phys. Rev. Lett.}, volume = {120}, pages = {121801}, year = {2018},
  doi = {10.1103/PhysRevLett.120.121801}, eprint = {1711.05623}, archivePrefix = {arXiv}
}

@article{LHCb2026,
  author = {Aaij, Roel and others}, collaboration = {LHCb},
  title = {Measurement of the Ratio of Branching Fractions {\(\mathcal B(B_c^+\to J/\psi\tau^+\nu_\tau)/\mathcal B(B_c^+\to J/\psi\mu^+\nu_\mu)\)}},
  journal = {Phys. Rev. Lett.}, volume = {137}, pages = {141801}, year = {2026},
  doi = {10.1103/13fb-1gc2}, eprint = {2606.10018}, archivePrefix = {arXiv}
}

@article{CMS2024,
  author = {Hayrapetyan, Aram and others}, collaboration = {CMS},
  title = {Test of lepton flavor universality in semileptonic {\(B_c^+\)} meson decays in proton-proton collisions at {\(\sqrt{s}=13\)} TeV},
  journal = {Phys. Rev. D}, volume = {111}, pages = {L051102}, year = {2025},
  doi = {10.1103/PhysRevD.111.L051102}, eprint = {2408.00678}, archivePrefix = {arXiv}
}

@article{CMS2025,
  author = {Chekhovsky, V. and others}, collaboration = {CMS},
  title = {Measurement of the ratio of the {\(B_c^+\to J/\psi\tau^+\nu_\tau\)} and {\(B_c^+\to J/\psi\mu^+\nu_\mu\)} branching fractions using three-prong {\(\tau\)} lepton decays},
  journal = {Phys. Rev. D}, volume = {113}, pages = {L111101}, year = {2026},
  doi = {10.1103/mw5c-5rw3}, eprint = {2510.21559}, archivePrefix = {arXiv}
}

@article{HPQCD2020,
  author = {Harrison, Judd and Davies, Christine T. H. and Lytle, Andrew},
  collaboration = {HPQCD}, title = {\(B_c\to J/\psi\) form factors for the full \(q^2\) range from lattice QCD},
  journal = {Phys. Rev. D}, volume = {102}, pages = {094518}, year = {2020},
  doi = {10.1103/PhysRevD.102.094518}, eprint = {2007.06957}, archivePrefix = {arXiv}
}

@article{HPQCD2025,
  author = {Harrison, Judd}, collaboration = {HPQCD},
  title = {Improved lattice {QCD} {\(B_c\to J/\psi\)} vector, axial-vector, and tensor form factors},
  journal = {Phys. Rev. D}, volume = {112}, pages = {034503}, year = {2025},
  doi = {10.1103/wll6-z4cb}, eprint = {2503.15090}, archivePrefix = {arXiv}
}

@article{HPQCDLFU2020,
  author = {Harrison, Judd and Davies, Christine T. H. and Lytle, Andrew},
  collaboration = {HPQCD},
  title = {\(R(J/\psi)\) and {\(B_c^-\to J/\psi\ell^-\bar\nu_\ell\)} Lepton Flavor Universality Violating Observables from Lattice {QCD}},
  journal = {Phys. Rev. Lett.}, volume = {125}, pages = {222003}, year = {2020},
  doi = {10.1103/PhysRevLett.125.222003}, eprint = {2007.06956}, archivePrefix = {arXiv}
}

@article{BBL,
  author = {Bodwin, Geoffrey T. and Braaten, Eric and Lepage, G. Peter},
  title = {Rigorous {QCD} analysis of inclusive annihilation and production of heavy quarkonium},
  journal = {Phys. Rev. D}, volume = {51}, pages = {1125--1171}, year = {1995},
  doi = {10.1103/PhysRevD.51.1125}, eprint = {hep-ph/9407339}, archivePrefix = {arXiv}
}

@article{PDG2026,
  author = {Takahashi, F. and others},
  collaboration = {Particle Data Group},
  title = {Review of Particle Physics},
  journal = {Int. J. Mod. Phys. A}, volume = {41}, pages = {2630011},
  year = {2026}, doi = {10.1142/S0217751X26300115}
}

@article{QiaoZhu,
  author = {Qiao, Cong-Feng and Zhu, Rui-Lin},
  title = {Estimation of semileptonic decays of {\(B_c\)} meson to {\(S\)}-wave charmonia with NRQCD},
  journal = {Phys. Rev. D}, volume = {87}, pages = {014009}, year = {2013},
  doi = {10.1103/PhysRevD.87.014009}, eprint = {1208.5916}, archivePrefix = {arXiv}
}

@article{ZhuRelativistic,
  author = {Zhu, Ruilin and Ma, Yan and Han, Xin-Ling and Xiao, Zhen-Jun},
  title = {Relativistic corrections to the form factors of {\(B_c\)} into {\(S\)}-wave charmonium},
  journal = {Phys. Rev. D}, volume = {95}, pages = {094012}, year = {2017},
  doi = {10.1103/PhysRevD.95.094012}, eprint = {1703.03875}, archivePrefix = {arXiv}
}

@article{MurphySoni2018,
  author = {Murphy, Christopher W. and Soni, Amarjit},
  title = {Model-independent determination of {\(B_c^+\to\eta_c\ell^+\nu\)} form factors},
  journal = {Phys. Rev. D}, volume = {98}, pages = {094026}, year = {2018},
  doi = {10.1103/PhysRevD.98.094026}, eprint = {1808.05932}, archivePrefix = {arXiv}
}

@article{HoangRuiz,
  author = {Hoang, Andr{\'e} H. and Ruiz-Femen{\'i}a, Pedro},
  title = {Heavy pair production currents with general quantum numbers in dimensionally regularized nonrelativistic {QCD}},
  journal = {Phys. Rev. D}, volume = {74}, pages = {114016}, year = {2006},
  doi = {10.1103/PhysRevD.74.114016}, eprint = {hep-ph/0609151}, archivePrefix = {arXiv}
}

@misc{Pan,
  author = {Pan, Jichen},
  title = {Direct renormalization of flavor-changing currents in {vNRQCD} for {\(B_c^{(*)}\)} decays},
  note = {accepted by Chinese Physics C}, year = {2026},
  url = {https://cpc.ihep.ac.cn/en/article/id/a06797d2-b64a-407c-be38-bb341c903495},
  eprint = {2608.21771}, archivePrefix = {arXiv}
}

@article{Nogueira,
  author = {Nogueira, Paulo}, title = {Automatic Feynman graph generation},
  journal = {J. Comput. Phys.}, volume = {105}, pages = {279--289}, year = {1993},
  doi = {10.1006/jcph.1993.1074}
}

@misc{FORM5,
  author = {Davies, J. and Kaneko, T. and Marinissen, C. and Ueda, T. and Vermaseren, J. A. M.},
  title = {{FORM} version 5.0},
  year = {2026}, eprint = {2601.19982}, archivePrefix = {arXiv}, primaryClass = {hep-ph}
}

@article{ChetyrkinTkachov,
  author = {Chetyrkin, K. G. and Tkachov, F. V.},
  title = {Integration by parts: The algorithm to calculate {\(\beta\)}-functions in 4 loops},
  journal = {Nucl. Phys. B}, volume = {192}, pages = {159--204}, year = {1981},
  doi = {10.1016/0550-3213(81)90199-1}
}

@article{Kira,
  author = {Maierh{\"o}fer, Philipp and Usovitsch, Johann and Uwer, Peter},
  title = {Kira---A Feynman integral reduction program},
  journal = {Comput. Phys. Commun.}, volume = {230}, pages = {99--112}, year = {2018},
  doi = {10.1016/j.cpc.2018.04.012}, eprint = {1705.05610}, archivePrefix = {arXiv}
}

@article{RemiddiDE,
  author = {Remiddi, Ettore},
  title = {Differential equations for Feynman graph amplitudes},
  journal = {Nuovo Cim. A}, volume = {110}, pages = {1435--1452}, year = {1997},
  doi = {10.1007/BF03185566}, eprint = {hep-th/9711188}, archivePrefix = {arXiv}
}

@article{AMFlow,
  author = {Liu, Xiao and Ma, Yan-Qing},
  title = {AMFlow: A Mathematica package for Feynman integrals computation via auxiliary mass flow},
  journal = {Comput. Phys. Commun.}, volume = {283}, pages = {108565}, year = {2023},
  doi = {10.1016/j.cpc.2022.108565}, eprint = {2201.11669}, archivePrefix = {arXiv}
}

@article{Larin,
  author = {Larin, S. A.}, title = {The renormalization of the axial anomaly in dimensional regularization},
  journal = {Phys. Lett. B}, volume = {303}, pages = {113--118}, year = {1993},
  doi = {10.1016/0370-2693(93)90053-K}, eprint = {hep-ph/9302240}, archivePrefix = {arXiv}
}

@article{DuganGrinstein,
  author = {Dugan, Michael J. and Grinstein, Benjam{\'i}n},
  title = {On the vanishing of evanescent operators},
  journal = {Phys. Lett. B}, volume = {256}, pages = {239--244}, year = {1991},
  doi = {10.1016/0370-2693(91)90680-O}
}

@article{BenekeSmirnov,
  author = {Beneke, M. and Smirnov, V. A.}, title = {Asymptotic expansion of Feynman integrals near threshold},
  journal = {Nucl. Phys. B}, volume = {522}, pages = {321--344}, year = {1998},
  doi = {10.1016/S0550-3213(98)00138-2}, eprint = {hep-ph/9711391}, archivePrefix = {arXiv}
}

@article{Jantzen,
  author = {Jantzen, Bernd}, title = {Foundation and generalization of the expansion by regions},
  journal = {JHEP}, volume = {2011}, number = {12}, pages = {076}, year = {2011},
  doi = {10.1007/JHEP12(2011)076}, eprint = {1111.2589}, archivePrefix = {arXiv}
}

@article{BodwinGarciaLee,
  author = {Bodwin, Geoffrey T. and {Garcia i Tormo}, Xavier and Lee, Jungil},
  title = {Factorization in exclusive quarkonium production},
  journal = {Phys. Rev. D}, volume = {81}, pages = {114014}, year = {2010},
  doi = {10.1103/PhysRevD.81.114014}, eprint = {1003.0061}, archivePrefix = {arXiv}
}

@article{HerrlichNierste,
  author = {Herrlich, Stefan and Nierste, Ulrich},
  title = {Evanescent operators, scheme dependences and double insertions},
  journal = {Nucl. Phys. B}, volume = {455}, pages = {39--58}, year = {1995},
  doi = {10.1016/0550-3213(95)00474-7}, eprint = {hep-ph/9412375}, archivePrefix = {arXiv}
}

@phdthesis{BellThesis,
  author = {Bell, Guido},
  title = {Higher order {QCD} corrections in exclusive charmless {$B$} decays},
  school = {Ludwig-Maximilians-Universit{\"a}t M{\"u}nchen},
  year = {2006}, eprint = {0705.3133}, archivePrefix = {arXiv}, primaryClass = {hep-ph}
}

@article{BellSoftOverlap,
  author = {Bell, Guido and B{\"o}er, Philipp and Feldmann, Thorsten and Horstmann, Dennis and Shtabovenko, Vladyslav},
  title = {{\(B_c\to\eta_c\)} form factors at large recoil: interplay of soft-quark and soft-gluon dynamics},
  journal = {JHEP}, volume = {2025}, number = {09}, pages = {098}, year = {2025},
  doi = {10.1007/JHEP09(2025)098}, eprint = {2412.14149}, archivePrefix = {arXiv}
}

@article{BellSCET,
  author = {Bell, Guido and B{\"o}er, Philipp and Feldmann, Thorsten and Horstmann, Dennis and Shtabovenko, Vladyslav},
  title = {{\(B_c\to\eta_c\)} form factors at large recoil: {SCET} analysis and a three-loop consistency check},
  journal = {JHEP}, volume = {2026}, number = {10}, pages = {050}, year = {2026},
  doi = {10.1007/JHEP10(2026)050}, eprint = {2605.21314}, archivePrefix = {arXiv}
}

@article{QiaoSunYuan2012,
  author = {Qiao, Cong-Feng and Sun, Peng and Yuan, Feng},
  title = {{NLO QCD} Corrections to {$B_c$}-to-Charmonium Form Factors},
  journal = {JHEP},
  volume = {2012},
  number = {08},
  pages = {087},
  year = {2012},
  doi = {10.1007/JHEP08(2012)087},
  eprint = {1103.2025},
  archivePrefix = {arXiv},
  primaryClass = {hep-ph}
}

@article{TaoTensor2022,
  author = {Tao, Wei and Xiao, Zhen-Jun and Zhu, Ruilin},
  title = {Next-to-leading order {QCD} calculation of {$B_c$} to charmonium tensor form factors},
  journal = {Phys. Rev. D},
  volume = {105},
  pages = {114026},
  year = {2022},
  doi = {10.1103/PhysRevD.105.114026},
  eprint = {2204.06385},
  archivePrefix = {arXiv},
  primaryClass = {hep-ph}
}

@misc{TaoEtAl2026NNLO,
  author = {Tao, Wei and Xiao, Zhen-Jun and Zhu, Ruilin and Chang, Qin},
  title = {Complete {NNLO QCD} Calculations of {$B_c^{(*)}\to J/\psi(\eta_c)$} Form Factors and Semileptonic Decays},
  year = {2026},
  eprint = {2611.34263v1},
  archivePrefix = {arXiv},
  primaryClass = {hep-ph}
}

@article{WangShenPion,
  author={Wang, Yu-Ming and Shen, Yue-Long},
  title={Subleading power corrections to the pion-photon transition form factor in {QCD}},
  journal={JHEP}, volume={2017}, number={12}, pages={037}, year={2017},
  eprint={1706.05680}, archivePrefix={arXiv}, doi={10.1007/JHEP12(2017)037}
}

@article{GaoPion,
  author={Gao, Jing and Huber, Tobias and Ji, Yao and Wang, Yu-Ming},
  title={Next-to-next-to-leading-order {QCD} prediction for the photon-pion form factor},
  journal={Phys. Rev. Lett.}, volume={128}, pages={062003}, year={2022},
  eprint={2106.01390}, archivePrefix={arXiv}, doi={10.1103/PhysRevLett.128.062003}
}

@article{JiPionEM,
  author = {Ji, Yao and Shi, Bo-Xuan and Wang, Jian and Wang, Ye-Fan and Wang, Yu-Ming and Yu, Hui-Xin},
  title = {Next-to-Next-to-Leading-Order {QCD} Prediction for the Pion Form Factor},
  journal = {Phys. Rev. Lett.}, volume = {134}, pages = {221901}, year = {2025},
  eprint = {2411.03658}, archivePrefix = {arXiv}, doi = {10.1103/yx51-lc3q}
}

@article{HuangNucleonFF,
  author = {Huang, Yong-Kang and Shi, Bo-Xuan and Wang, Yu-Ming and Zhao, Xue-Chen},
  title = {Next-to-Leading-Order {QCD} Predictions for the Nucleon Form Factors},
  journal = {Phys. Rev. Lett.}, volume = {135}, pages = {061901}, year = {2025},
  eprint = {2407.18724}, archivePrefix = {arXiv}, doi = {10.1103/xz46-8ptk}
}

@article{ShiSigmaFF,
  author = {Shi, Bo-Xuan and Yu, Hui-Xin and Zhao, Xue-Chen},
  title = {Next-to-leading-order {QCD} predictions for the {$\Sigma$} Dirac form factors},
  journal = {Chin. Phys. C}, volume = {50}, pages = {053111}, year = {2026},
  eprint = {2602.13716}, archivePrefix = {arXiv}, doi = {10.1088/1674-1137/ae4ba9}
}

@misc{HuangNucleonDA,
  author = {Huang, Yong-Kang and Ji, Yao and Shi, Bo-Xuan and Wang, Yu-Ming},
  title = {Two-Loop Renormalization-Group Evolution for the Nucleon Distribution Amplitude},
  year = {2025}, eprint = {2512.20471}, archivePrefix = {arXiv}, primaryClass = {hep-ph}
}

@article{CaswellLepage,
  author = {Caswell, William E. and Lepage, G. Peter},
  title = {Effective lagrangians for bound state problems in {QED}, {QCD}, and other field theories},
  journal = {Phys. Lett. B}, volume = {167}, pages = {437--442}, year = {1986},
  doi = {10.1016/0370-2693(86)91297-9}
}

@article{CzarneckiMelnikovThreshold,
  author = {Czarnecki, Andrzej and Melnikov, Kirill},
  title = {Two-loop {QCD} corrections to the heavy quark pair production cross section in {\(e^+e^-\)} annihilation near the threshold},
  journal = {Phys. Rev. Lett.}, volume = {80}, pages = {2531--2534}, year = {1998},
  doi = {10.1103/PhysRevLett.80.2531}, eprint = {hep-ph/9712222}, archivePrefix = {arXiv}
}

@article{OnishchenkoVeretin,
  author = {Onishchenko, Andrei I. and Veretin, Oleg L.},
  title = {Two-loop {QCD} corrections to the leptonic constant of the {\(B_c\)} meson},
  journal = {Eur. Phys. J. C}, volume = {50}, pages = {801--808}, year = {2007},
  doi = {10.1140/epjc/s10052-007-0255-1}, eprint = {hep-ph/0302132}, archivePrefix = {arXiv}
}

@article{ChenQiaoLeptonic,
  author = {Chen, Long-Bin and Qiao, Cong-Feng},
  title = {Two-loop {QCD} corrections to {\(B_c\)} meson leptonic decays},
  journal = {Phys. Lett. B}, volume = {748}, pages = {443--450}, year = {2015},
  doi = {10.1016/j.physletb.2015.07.043}, eprint = {1503.05122}, archivePrefix = {arXiv}
}

@article{BroadhurstGraySchilcher,
  author = {Broadhurst, D. J. and Gray, N. and Schilcher, K.},
  title = {Gauge-invariant on-shell {\(Z_2\)} in {QED}, {QCD} and the effective field theory of a static quark},
  journal = {Z. Phys. C}, volume = {52}, pages = {111--122}, year = {1991},
  doi = {10.1007/BF01412333}
}

@article{BekavacOnShell,
  author = {Bekavac, S. and Grozin, A. and Seidel, D. and Steinhauser, M.},
  title = {Light quark mass effects in the on-shell renormalization constants},
  journal = {JHEP}, volume = {2007}, number = {10}, pages = {006}, year = {2007},
  doi = {10.1088/1126-6708/2007/10/006}, eprint = {0708.1729}, archivePrefix = {arXiv}
}

@article{GraceyTensor,
  author = {Gracey, J. A.},
  title = {Three loop {\(\overline{\mathrm{MS}}\)} tensor current anomalous dimension in {QCD}},
  journal = {Phys. Lett. B}, volume = {488}, pages = {175--181}, year = {2000},
  doi = {10.1016/S0370-2693(00)00859-5}, eprint = {hep-ph/0007171}, archivePrefix = {arXiv}
}

@article{TkachovIBP,
  author = {Tkachov, F. V.},
  title = {A theorem on analytical calculability of 4-loop renormalization group functions},
  journal = {Phys. Lett. B}, volume = {100}, pages = {65--68}, year = {1981},
  doi = {10.1016/0370-2693(81)90288-4}
}

@article{Laporta,
  author = {Laporta, S.},
  title = {High-precision calculation of multiloop {Feynman} integrals by difference equations},
  journal = {Int. J. Mod. Phys. A}, volume = {15}, pages = {5087--5159}, year = {2000},
  doi = {10.1142/S0217751X00002157}, eprint = {hep-ph/0102033}, archivePrefix = {arXiv}
}

@article{Kira2,
  author = {Klappert, Jonas and Lange, Fabian and Maierh{\"o}fer, Philipp and Usovitsch, Johann},
  title = {Integral reduction with {Kira 2.0} and finite field methods},
  journal = {Comput. Phys. Commun.}, volume = {266}, pages = {108024}, year = {2021},
  doi = {10.1016/j.cpc.2021.108024}, eprint = {2008.06494}, archivePrefix = {arXiv}
}

@article{Kira3,
  author = {Lange, Fabian and Usovitsch, Johann and Wu, Zihao},
  title = {{Kira 3}: integral reduction with efficient seeding and optimized equation selection},
  journal = {Comput. Phys. Commun.}, volume = {322}, pages = {109999}, year = {2026},
  doi = {10.1016/j.cpc.2025.109999}, eprint = {2505.20197}, archivePrefix = {arXiv}
}

@article{KotikovDE,
  author = {Kotikov, A. V.},
  title = {Differential equations method. New technique for massive {Feynman} diagram calculation},
  journal = {Phys. Lett. B}, volume = {254}, pages = {158--164}, year = {1991},
  doi = {10.1016/0370-2693(91)90413-K}
}

@article{GehrmannRemiddiDE,
  author = {Gehrmann, Thomas and Remiddi, Ettore},
  title = {Differential equations for two-loop four-point functions},
  journal = {Nucl. Phys. B}, volume = {580}, pages = {485--518}, year = {2000},
  doi = {10.1016/S0550-3213(00)00223-6}, eprint = {hep-ph/9912329}, archivePrefix = {arXiv}
}

@article{BinothHeinrich,
  author = {Binoth, T. and Heinrich, G.},
  title = {An automatized algorithm to compute infrared divergent multi-loop integrals},
  journal = {Nucl. Phys. B}, volume = {585}, pages = {741--759}, year = {2000},
  doi = {10.1016/S0550-3213(00)00429-6}, eprint = {hep-ph/0004013}, archivePrefix = {arXiv}
}

@article{KanekoUeda,
  author = {Kaneko, Toshiaki and Ueda, Takahiro},
  title = {A geometric method of sector decomposition},
  journal = {Comput. Phys. Commun.}, volume = {181}, pages = {1352--1361}, year = {2010},
  doi = {10.1016/j.cpc.2010.04.001}, eprint = {0908.2897}, archivePrefix = {arXiv}
}

@article{pySecDec,
  author = {Borowka, S. and Heinrich, G. and Jahn, S. and Jones, S. P. and Kerner, M. and Schlenk, J. and Zirke, T.},
  title = {{pySecDec}: a toolbox for the numerical evaluation of multi-scale integrals},
  journal = {Comput. Phys. Commun.}, volume = {222}, pages = {313--326}, year = {2018},
  doi = {10.1016/j.cpc.2017.09.015}, eprint = {1703.09692}, archivePrefix = {arXiv}
}

@article{MorielloDE,
  author = {Moriello, Francesco},
  title = {Generalised power series expansions for the elliptic planar families of {Higgs} + jet production at two loops},
  journal = {JHEP}, volume = {2020}, number = {01}, pages = {150}, year = {2020},
  doi = {10.1007/JHEP01(2020)150}, eprint = {1907.13234}, archivePrefix = {arXiv}
}

@article{LiuMaWangAMF,
  author = {Liu, Xiao and Ma, Yan-Qing and Wang, Chen-Yu},
  title = {A systematic and efficient method to compute multi-loop master integrals},
  journal = {Phys. Lett. B}, volume = {779}, pages = {353--357}, year = {2018},
  doi = {10.1016/j.physletb.2018.02.026}, eprint = {1711.09572}, archivePrefix = {arXiv}
}

@article{LiuMaAMF,
  author = {Liu, Xiao and Ma, Yan-Qing},
  title = {Multiloop corrections for collider processes using auxiliary mass flow},
  journal = {Phys. Rev. D}, volume = {105}, pages = {L051503}, year = {2022},
  doi = {10.1103/PhysRevD.105.L051503}, eprint = {2107.01864}, archivePrefix = {arXiv}
}

@article{LiuMaIteratedAMF,
  author = {Liu, Zhi-Feng and Ma, Yan-Qing},
  title = {Determining {Feynman} integrals with only input from linear algebra},
  journal = {Phys. Rev. Lett.}, volume = {129}, pages = {222001}, year = {2022},
  doi = {10.1103/PhysRevLett.129.222001}, eprint = {2201.11637}, archivePrefix = {arXiv}
}

@article{LiuMaTaoZhangAMF,
  author = {Liu, Xiao and Ma, Yan-Qing and Tao, Wei and Zhang, Peng},
  title = {Calculation of {Feynman} loop integration and phase-space integration via auxiliary mass flow},
  journal = {Chin. Phys. C}, volume = {45}, pages = {013115}, year = {2021},
  doi = {10.1088/1674-1137/abc538}, eprint = {2009.07987}, archivePrefix = {arXiv}
}

@misc{AMFlow2,
  author = {Huang, Rui-Jun and Liu, Xiao and Ma, Yan-Qing},
  title = {{AMFlow 2.0}: significant algorithmic and software improvements for {Feynman} integral evaluation},
  year = {2026}, eprint = {2607.08477}, archivePrefix = {arXiv}, primaryClass = {hep-ph}
}

@article{ChenDoubleCharmonium,
  author = {Chen, Xiang and Guan, Xin and He, Chuan-Qi and Ma, Yan-Qing and Wang, Jian and Zhang, Da-Jiang},
  title = {Analytical two-loop amplitudes of {\(e^+e^-\to J/\psi+\eta_c\)} at {\(B\)} factories},
  journal = {Phys. Rev. D}, volume = {113}, pages = {074023}, year = {2026},
  doi = {10.1103/93kb-dqz6}, eprint = {2508.20777}, archivePrefix = {arXiv}, primaryClass = {hep-ph}
}

@article{CaswellBeta,
  author = {Caswell, William E.},
  title = {Asymptotic behavior of non-{Abelian} gauge theories to two-loop order},
  journal = {Phys. Rev. Lett.}, volume = {33}, pages = {244--246}, year = {1974},
  doi = {10.1103/PhysRevLett.33.244}
}

@article{JonesBeta,
  author = {Jones, D. R. T.},
  title = {Two-loop diagrams in {Yang--Mills} theory},
  journal = {Nucl. Phys. B}, volume = {75}, pages = {531--538}, year = {1974},
  doi = {10.1016/0550-3213(74)90093-5}
}

@article{BordoneSumRules,
  author = {Bordone, Marzia and Khodjamirian, Alexander and Mannel, Thomas},
  title = {New sum rules for the {\(B_c\to J/\psi\)} form factors},
  journal = {JHEP}, volume = {2023}, number = {01}, pages = {032}, year = {2023},
  doi = {10.1007/JHEP01(2023)032}, eprint = {2209.08851}, archivePrefix = {arXiv}
}

@article{TranIvanovKornerSantorelli,
  author = {Tran, Chien-Thang and Ivanov, Mikhail A. and K{\"o}rner, J{\"u}rgen G. and Santorelli, Pietro},
  title = {Implications of new physics in the decays {\(B_c\to(J/\psi,\eta_c)\tau\nu\)}},
  journal = {Phys. Rev. D}, volume = {97}, pages = {054014}, year = {2018},
  doi = {10.1103/PhysRevD.97.054014}, eprint = {1801.06927}, archivePrefix = {arXiv}
}

\end{document}


\setlength{\abovedisplayskip}{8pt plus 2pt minus 2pt}
\setlength{\belowdisplayskip}{\abovedisplayskip}
\setcounterref{equation}{main-eq:alt}
\begin{center}
 {\LARGE\bfseries Supplemental Material\par}
\end{center}

\section{Born form factors}
\label{sec:s_anc}

Let \(y=q^2/m_b^2\). In the wave-function convention of
Eq.~(\ref*{main-eq:meson-currents}), the Born form factors share the
normalization
\begin{equation}
 {\cal N}_H=\frac{8\pi\alpha_s(\mu_R) C_F\sqrt{2(1+z)}\,
                  \psi_{B_c}(0)\psi_H^*(0)}
                 {z^{3/2}m_b^3[(1-z)^2-y]^2}.
 \label{eq:s_born-normalization}
\end{equation}
The Born expressions are~\cite{QiaoZhu,QiaoSunYuan2012,TaoTensor2022}
\begin{align}
 (V,A_0,A_1,A_2)^{(0)}&=2{\cal N}_{J/\psi}
 \Bigl((1+3z)(1+z),\ (1+z)^2,\
 \frac{4z^3+5z^2+6z+1-y(2z+1)}{1+3z},\ 1+3z\Bigr),\nonumber\\
 (f_+,f_0,f_T)^{(0)}&={\cal N}_{\eta_c}
 \left(b_+,\ b_+-\frac{2y(1-z^2)}{(1+z)^2-4z^2},\
                   2(1+z)(1+3z)\right).
 \label{eq:s_born-factors}
\end{align}
Here \(b_+=3+2z+3z^2-y\). We check the normalization of \(V\) and \(f_+\) by a direct
tree-level calculation with
\(f_{\rm meson}=\sqrt{4N_c/M_{\rm meson}}\,\psi(0)\).

\section{Evanescent operators and two-loop mixing}
\label{sec:s_evanescent}

We give all Born evanescent operators and their finite mixing
coefficients \(Z_{EP}^{(2)}\), expressed in terms of
\(g(r)=4\pi^2 C_F^2r/[3(1+r)^2]\).
The symbol \(\varepsilon^{\parallel ab}\) is the three-dimensional
Levi-Civita symbol with \(\varepsilon^{\parallel12}=1\) and vanishes
if either index lies outside the physical transverse plane.

For \(\eta_c\), the scalar and longitudinal vector currents and the
tensor contraction with \(v_\mu v'_\nu\) share the evanescent operator
\begin{equation}
 \begin{aligned}[b]
 E_\eta
 &= (\psi_c^\dagger\psi_b)(\chi_c^\dagger\chi_c)
   +\tfrac12 P_{\eta_c}
   +\tfrac12\sum_i
       (\psi_c^\dagger\Sigma^i\chi_c)(\chi_c^\dagger\Sigma^i\psi_b).
 \end{aligned}
 \label{eq:s_evanescent_eta}
\end{equation}
Its mixing coefficient is \(Z_{E_\eta P_{\eta_c}}^{(2)}=5(g(z)+g(1))\).
The scalar current also contains \(E_\perp\) of
Eq.~(\ref*{main-eq:evanescent-line}), with the mixing coefficient
in Eq.~(\ref*{main-eq:born-ep-mixing}).

For \(J/\psi\), the transverse vector current contains
\begin{equation}
 \begin{aligned}[b]
 E_V^a
 &= (\psi_c^\dagger\Sigma^\parallel\Sigma^a\psi_b)
    (\chi_c^\dagger\chi_c)\\
 &\quad+\tfrac12\big[
       (\psi_c^\dagger\Sigma^\parallel\chi_c)
       (\chi_c^\dagger\Sigma^a\psi_b)
       -(\psi_c^\dagger\Sigma^a\chi_c)
       (\chi_c^\dagger\Sigma^\parallel\psi_b)\big]\\
 &\quad+\tfrac{i}{2}\sum_b\varepsilon^{\parallel ab}
       \big[P_{J/\psi}^b+(\psi_c^\dagger\chi_c)
       (\chi_c^\dagger\Sigma^b\psi_b)\big].
 \end{aligned}
 \label{eq:s_evanescent_vector}
\end{equation}
Its mixing coefficient is
\(Z_{E_V^a P_{J/\psi}^b}^{(2)}=i\varepsilon^{\parallel ab}(5g(z)+g(1))\).

The longitudinal axial current contains two evanescent operators,
\begin{equation}
 \begin{aligned}[b]
 E_{A_\parallel,1}
 &= (\psi_c^\dagger\Sigma^\parallel\psi_b)(\chi_c^\dagger\chi_c)
    +\tfrac12\big[P_{J/\psi}^\parallel+
      (\psi_c^\dagger\chi_c)(\chi_c^\dagger\Sigma^\parallel\psi_b)\big]\\
 &\quad-\tfrac{i}{2}\sum_{a,b}\varepsilon^{\parallel ab}
      (\psi_c^\dagger\Sigma^a\chi_c)(\chi_c^\dagger\Sigma^b\psi_b),\\
 E_{A_\parallel,2}
 &= \sum_b(\psi_c^\dagger\Sigma^b\psi_b)
       (\chi_c^\dagger\Sigma^b\Sigma^\parallel\chi_c)
       +P_{J/\psi}^\parallel
       -(\psi_c^\dagger\chi_c)(\chi_c^\dagger\Sigma^\parallel\psi_b).
 \end{aligned}
 \label{eq:s_evanescent_axial_longitudinal}
\end{equation}
They mix into \(P_{J/\psi}^{\parallel}\) with coefficients \(5g(z)+g(1)\)
and \(4g(z)\), respectively. The pseudoscalar current also contains
\(E_{A_\parallel,1}\).

The transverse axial current involves
\begin{equation}
 \begin{aligned}[b]
 E_{A_\perp,1}^a
 &= (\psi_c^\dagger\Sigma^\parallel\psi_b)
      (\chi_c^\dagger\Sigma^a\Sigma^\parallel\chi_c)
      +\tfrac12\big[(\psi_c^\dagger\chi_c)
      (\chi_c^\dagger\Sigma^a\psi_b)-P_{J/\psi}^a\big]\\
 &\quad-\tfrac{i}{2}\sum_b\varepsilon^{\parallel ab}\big[
      (\psi_c^\dagger\Sigma^\parallel\chi_c)
      (\chi_c^\dagger\Sigma^b\psi_b)
      +(\psi_c^\dagger\Sigma^b\chi_c)
      (\chi_c^\dagger\Sigma^\parallel\psi_b)\big],\\
 E_{A_\perp,2}^a
 &= (\psi_c^\dagger\Sigma^a\psi_b)(\chi_c^\dagger\chi_c)
       +\tfrac12\big[(\psi_c^\dagger\chi_c)
      (\chi_c^\dagger\Sigma^a\psi_b)+P_{J/\psi}^a\big]\\
 &\quad+\tfrac{i}{2}\sum_b\varepsilon^{\parallel ab}\big[
      (\psi_c^\dagger\Sigma^\parallel\chi_c)
      (\chi_c^\dagger\Sigma^b\psi_b)
      -(\psi_c^\dagger\Sigma^b\chi_c)
      (\chi_c^\dagger\Sigma^\parallel\psi_b)\big],\\
 E_{A_\perp,3}^a
 &= \sum_{b\ne a}(\psi_c^\dagger\Sigma^\parallel\Sigma^a\Sigma^b\psi_b)
       (\chi_c^\dagger\Sigma^b\Sigma^\parallel\chi_c)
      -\tfrac12\big[(\psi_c^\dagger\chi_c)
      (\chi_c^\dagger\Sigma^a\psi_b)+P_{J/\psi}^a\big]\\
 &\quad+\tfrac{i}{2}\sum_b\varepsilon^{\parallel ab}\big[
      (\psi_c^\dagger\Sigma^\parallel\chi_c)
      (\chi_c^\dagger\Sigma^b\psi_b)
      -(\psi_c^\dagger\Sigma^b\chi_c)
      (\chi_c^\dagger\Sigma^\parallel\psi_b)\big].
 \end{aligned}
 \label{eq:s_evanescent_axial_transverse}
\end{equation}
For each \(a\), the finite mixing coefficients are
\[
 Z_{E_{A_\perp,1}^a P_{J/\psi}^a}^{(2)}=-5g(z)-g(1),\qquad
 Z_{E_{A_\perp,2}^a P_{J/\psi}^a}^{(2)}=5g(z)+g(1),\qquad
 Z_{E_{A_\perp,3}^a P_{J/\psi}^a}^{(2)}=g(z)+g(1).
\]

\renewcommand{\bibfont}{\small}\setlength{\bibsep}{0pt}
\renewcommand{\bibitemNoStop}{\unskip.\spacefactor=3000\space}
\bibliographystyle{apsrev4-2}
\bibliography{references}